\documentclass[fleqn,twoside]{article}
\usepackage{epsfig}
\usepackage{graphicx}
\usepackage{amssymb}
\usepackage{here}
\usepackage{espcrc2}
\newcommand{\AmS}{{\protect\the\textfont2
  A\kern-.1667em\lower.5ex\hbox{M}\kern-.125emS}}

\title{\center{\large{\bf Hunting a light CP-odd non-standard Higgs boson 
through its tauonic decay \\ 
at a (Super) B factory}}
\thanks{Report:IFIC/07-11,FTUV-07-0217; 
Research under grants: FPA-2003-09229-C01, 
FPA2005-01678 and GVACOMP2006}}
\author{\center{Esteban Fullana$^a$ and 
Miguel-Angel Sanchis-Lozano$^{a,b}$\thanks{Email:Miguel.Angel.Sanchis@uv.es}
\vspace{0.1cm}\\
\it $^a$Instituto de F\'{\i}sica
Corpuscular (IFIC) and $^b$Departamento de F\'{\i}sica Te\'orica \\
\it Centro Mixto Universidad de Valencia-CSIC, 
Dr. Moliner 50, E-46100 Burjassot, Valencia (Spain)}}
\begin{document}

\begin{abstract}

Several scenarios beyond the minimal extension of the Standard Model still
allow light non-standard Higgs bosons evading LEP bounds. We examine
the mixing between a light CP-odd Higgs boson and $\eta_b$ states
and its implications on a slight (but observable) lepton
universality breaking in Upsilon decays 
to be measured at the percent level at a (Super) B factory.  
\end{abstract}

\maketitle

\section{Introduction}
With the advent of the LHC era, the search for signals
of New Physics (NP) beyond the
Standard Model (SM) is definitely becoming one of the hottest topics
in elementary particle physics. On the other hand, it is widely 
recognized that the high-energy frontier of discovery has
to be complemented by high-precision experiments
at lower energies - like those performed
at high luminosity B factories -  
providing valuable information for 
the LHC and ILC themselves.

The relevance of (radiative) decays of the $\Upsilon$ resonance
in the quest for (pseudo)scalar non-standard particles was soon
recognized after its discovery 
\cite{Wilczek:1977pj,Haber:1978jt,Ellis:1979jy}, quickly followed by
experimental searches
\cite{Peck:1984vx,Besson:1985xw,Albrecht:1985qz} which, however, 
have yielded negative results so far. (No further confirmation 
was found for a narrow state claimed by Crystal Ball with a mass
around 8.3 GeV.) Basically, in all these searches a monochromatic
photon was expected (for a narrow Higgs boson)
but no peak was observed in the photon
spectrum and, therefore, narrow intermediate 
states were excluded in the analysis. 

As argued in this paper, the mixing between $^1S_0$ bottomonium states
and a light CP-odd Higgs boson ($A^0$) would
imply broader intermediate states, e.g. 
in $\Upsilon \to \gamma\ \eta_b(\to \tau^+\tau^-)$ decays, 
yielding rather non-monochromatic radiated photons. Thus, any
photon signal peak would be smeared and probably swallowed up in the
background rendering difficult the experimental observation
in this way. Moreover, the photon energy region $\lesssim$ 100 MeV
in $\Upsilon(1S) \to \gamma+X$ decays
is generally not examined 
by experiments \cite{Yao:2006px} because of a low photon
detection efficiency.
Nevertheless, NP might still show up as a
(slight) breaking of lepton universality (LU) 
in $\Upsilon$ decays as advocated in  
\cite{Sanchis-Lozano:2006gx,Sanchis-Lozano:2003ha}.

From a theoretical viewpoint, the existence of a light
pseudoscalar Higgs is not unexpected in certain non-minimal
extensions of the SM. As an especially appealing example, the 
next-to-minimal supersymmetric standard 
model (NMSSM)  gets a gauge singlet added to the
MSSM two-doublet Higgs sector, leading
to seven physical Higgs bosons, five of them neutral including
two pseudoscalars \cite{gunion}. In the limit of either
slightly broken $R$ or Peccei-Quinn symmetries, the lightest
CP-odd Higgs would be naturally light
\cite{Dobrescu:2000yn,Hiller:2004ii}, even requiring
the smallest degree of fine tuning \cite{Dermisek:2005ar}.

Interestingly, the authors of \cite{Dermisek:2005gg}
interpret, within the NMSSM, the excess of $Z$+$b$-jets events found at LEP  
as a signal of a 
SM-like Higgs decaying partly into $b\bar{b}\tau^+\tau^-$, but
dominantly into $\tau$'s via two light pseudoscalars.
Let us also mention the exciting connection with
possible light neutralino dark matter  
\cite{Gunion:2005rw} and its detection
at B factories \cite{McElrath:2005bp}. 
In Ref.~\cite{Dermisek:2006py}
a thorough analysis of the $\Upsilon$ radiative decay
into a pseudoscalar Higgs can be found, except for 
a mass range close to the resonance which is the main object of the
present study.

The possibility of light Higgs particles can be extended to 
scenarios with more than one gauge singlet \cite{Han:2004yd}, 
and even to the MSSM with a CP-violating Higgs sector 
\cite{Carena:2002bb} as 
LEP bounds can be then evaded  \cite{Kraml:2006ga}.
Indeed, in the CP-violating benchmark scenario and several variants,  
the combined LEP data show large domains of the parameter space
which are not excluded, down to the lowest Higgs mass values  
\cite{Schael:2006cr}. A similar conclusion applies to  
a two Higgs doublet model of type II (2HDM(II)) \cite{gunion}, where 
some windows for a very light Higgs are still open
\cite{Abbiendi:2004gn}.\\

In addition, Little Higgs models have an extended structure 
of global symmetries
(among which there can appear $U(1)$ factors) broken
both spontaneously and explicitly, leading to possible light pseudoscalar
particles in the Higgs spectrum.
The mass of such pseudoaxions is, in fact, not predicted by
the model and small values (of order of few GeV)
are allowed \cite{Kraml:2006ga}.
Finally, let us mention the $g-2$ muon anomaly \cite{Bennett:2006fi}, 
which might require
a light CP-odd Higgs boson \cite{Krawczyk:2001pe}
to reconcile the experimental value with the SM result
\cite{Hagiwara:2006jt}. 

In this investigation we consider $^3S_1$ bottomonium states
below the $B\bar{B}$ threshold
undergoing a radiative decay yielding a 
light non-standard Higgs boson denoted by $A^0$. The latter 
particle can decay
into a lepton pair with a sizeable branching fraction (BF) if
its coupling to fermions is enhanced enough, 
according to some scenarios beyond the SM.
As emphasized in 
\cite{Sanchis-Lozano:2003ha,Sanchis-Lozano:2005di}, 
the NP contribution would be unwittingly ascribed to the 
$\Upsilon$ tauonic channel thereby breaking LU
if the (not necessarily soft) radiated photon 
escapes undetected in the experiment \footnote{Note that
the leptonic width as expressed in Ref.\cite{Yao:2006px} is, in fact, 
an inclusive quantity with a sum over an infinite number of photons}.
Let us remark that higher-order decays within the SM
breaking LU 
(like $Z^0$-exchange 
or two-photon (one-loop) annihilation of intermediate $\eta_b$ states
into leptons \cite{Haber:1978jt}) are negligible.

Basically, two radiative decay modes of
Upsilon resonances will be considered 
in our analysis:
\begin{itemize}
\item[$i)$] {\em Non-resonant decay:} $\Upsilon$ resonances can 
(directly) decay by emitting a photon yielding a final-state $A^0$,
which subsequently decays into a $\tau^+\tau^-$ pair
as shown in Fig.1.a), 
\begin{equation}
\Upsilon(nS)\ {\rightarrow}\ {\bf \gamma}\ 
A^0 ({\rightarrow}\ \tau^+\tau^-)
\end{equation}
\item[$ii)$] {\em Resonant decay:} This mechanism corresponds to
the process depicted in figure 1.b),
\begin{equation}
\Upsilon(nS) {\rightarrow}\ {\bf \gamma}\
\eta_b(n'S) (\rightarrow
A^{0*} {\rightarrow}\ \tau^+\tau^-)
\end{equation}
\end{itemize}
When neglecting bound-state effects, the process of Eq.(1) corresponds to the
leading-order Wilczek mechanism \cite{Wilczek:1977pj}.
In section 4 we shall take into account 
the mixing between the $\eta_b$ and the pseudoscalar boson $A^0$ 
generated by the diagram 1.c), 
following the guidelines of Ref.~\cite{Drees:1989du}, 
but updating several inputs from
the bottomonium and Higgs sectors.

\begin{figure}
\begin{center}
\includegraphics[width=13pc]{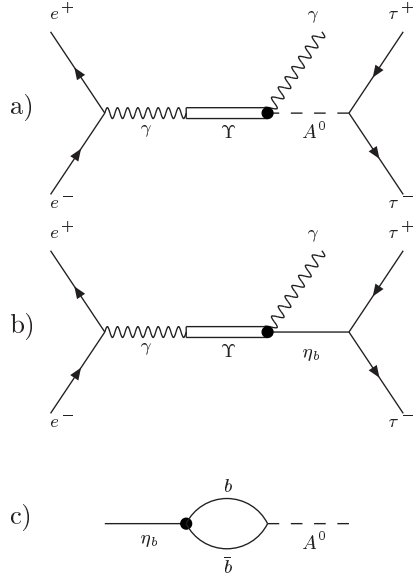}
\end{center}
\caption{Contributions to $e^+e^-\to \Upsilon \to \gamma\ \tau^+\tau^-$ 
with a) pseudoscalar Higgs, b) $\eta_b$, as
intermediate states; c) Mixing diagram to be considered
in section 4.}
\end{figure}

For our theoretical estimates we
assume that fermions couple to the $A^0$ field
according to the interaction term
\begin{equation}
{\cal L}_{int}^{\bar{f}f}\ =\ -X_f\ \frac{A^0}{v}
m_f\bar{f}(i\gamma_5)f
\label{eq:yukawa}
\end{equation}
in the effective Lagrangian, with the vacuum expectation value $v
\simeq 246$ GeV; $X_f$ depends on the fermion type, whose
mass is denoted by $m_f$.

\begin{table*}[hbt]
\setlength{\tabcolsep}{0.4pc}
\caption{Measured leptonic branching fractions
${\mathcal{B}[\Upsilon(nS) \to \ell \ell]}$ (in $\%$) and error bars
(summed in quadrature) of
$\Upsilon(1S)$, $\Upsilon(2S)$ and $\Upsilon(3S)$ resonances
(obtained from the PDG \cite{Yao:2006px} and recent CLEO data
\cite{Besson:2006gj}).\newline}
\label{FACTORES}

\begin{center}
\begin{tabular}{ccccc}
\hline
channel: & $e^+e^-$ & $\mu^+\mu^-$ & $\tau^+\tau^-$ &
$R_{\tau/\ell}(nS)$ \\
\hline
$\Upsilon(1S)$ & $2.38 \pm 0.11$ &  & $2.61 \pm 0.13$ &
$0.10 \pm 0.07$\\
\hline
$\Upsilon(1S)$ &             & $2.48 \pm 0.05$ & $2.61 \pm 0.13$ &
$0.05 \pm 0.06$\\
\hline
$\Upsilon(2S)$ & $1.97 \pm 0.11$ &    & $2.11 \pm 0.15$ &
$0.07 \pm 0.09$ \\
\hline
$\Upsilon(2S)$ & & $1.93 \pm 0.17$ & $2.11 \pm 0.15$ &
$0.09 \pm 0.12$ \\
\hline
$\Upsilon(3S)$ & $2.18 \pm 0.20$ &    & $2.52 \pm 0.24$ &
$0.14 \pm 0.14$ \\
\hline
$\Upsilon(3S)$ & & $2.18 \pm 0.21$ & $2.52 \pm 0.24$ &
$0.14 \pm 0.14$ \\
\hline
\end{tabular}
\end{center}
\end{table*}

\begin{figure}
\begin{center}
\includegraphics[width=18pc]{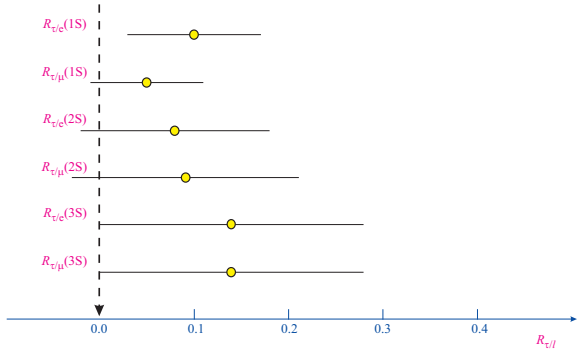}
\end{center}
\caption{Plot of $R_{\tau/\ell}$ according to Table 1.
A small but systematic shift towards positive values can be observed
which, however, needs confirmation.}
\end{figure}

In a 2HDM(II), $X_d=\tan{\beta}$ for down-type
fermions (where $\tan{\beta}$ stands for the ratio of two Higgs
doublet vacuum expectation values) and 
$X_u=\cot{\beta}$ for
up-type fermions. Large $\tan{\beta}$ values would imply a
large coupling of the $A^0$ to the bottom quark but a small
coupling to the charm quark, so hereafter we focus on the Upsilon 
phenomenology. 

In the NMSSM, 
$X_d=\cos{\theta_A} \tan{\beta}$ where
$\theta_A$ is the mixing angle between the singlet component
(which does not couple to $b\bar{b}$)
and the MSSM-like component of the $A^0$:
$A^0=\cos{\theta_A}A_{MSSM}+\sin{\theta_A}A_s$. Notice
that for large $\tan{\beta}$, $\cos{\theta_A} \sim \sin{2\beta}$
and no further increase of the coupling with $\tan{\beta}$
is expected in this limit. Thus, the constraints on a light 
CP-odd NMSSM Higgs boson from direct
searches at LEP are rather weak.

Experimentally, the relative importance of the postulated NP
contribution can be assessed via the ratio
of leptonic BF's
\begin{equation}
{\cal R}_{\tau/\ell}= \frac{{\cal B}_{\tau\tau}-{\cal B}_{\ell\ell}}
{{\cal B}_{\ell\ell}}= \frac{{\cal B}_{\tau\tau}}{{\cal B}_{\ell\ell}}-1
\label{eq:R}
\end{equation}
where ${\cal B}_{\tau\tau}$ denotes the tauonic BF 
of the $\Upsilon$ resonance and
${\cal B}_{\ell\ell}$, $\ell=e,\mu$ represents
either its electronic or muonic BF. 

A (statistically significant) non-null value of
${\cal R}_{\tau/\ell}$ would imply the rejection
of LU (predicting ${\cal R}_{\tau/\ell} \simeq 0$)
and a strong argument
supporting the existence of a pseudoscalar Higgs boson mediating
the tauonic channel according to the channels of Eqs.(1)
and (2). Table 1 shows current experimental data 
for the $\Upsilon(1S,2S,3S)$
resonances, pointing out the possibility of LU
breaking at the $\lesssim 10\%$ level. 

\section{Non-resonant decay} 
Soon after the discovery of heavy vector mesons,
Wilczek \cite{Wilczek:1977pj} put forward 
the exciting possibility of finding
a light Higgs or an axion through the radiative decay
$\Upsilon \to \gamma + H^0,A^0$.
The (pseudo)scalar particle could leave the detector volume without
decaying inside, or decaying mainly into charmed particles
or $\tau^+\tau^-$ pairs. 
In either case, an experimental signature was expected to be
a rather clean peak in the photon spectrum \cite{Albrecht:1985qz}
due to a foreseen small Higgs boson width.

The leading-order expression for the decay width 
(common to CP-even and -odd particles) reads
\[
\Gamma[\Upsilon(nS) \to \gamma A^0]=
\frac{2\alpha Q_b^2|R_n(0)|^2 X_d^2}
{v^2}\biggl(1-\frac{m_{A^0}^2}{m_{\Upsilon}^2}\biggr)
\]
for $m_{A^0} < m_{\Upsilon(nS)}$, where $R_n(0)$ and $m_{\Upsilon}$ denote 
the radial wave function at the origin
and mass of the $\Upsilon(nS)$, respectively;    
$\alpha$ is the fine structure constant and $Q_b$ 
is the electric charge (in units of $|e|$) of the bottom quark.
The above width is often divided by the electronic partial width
$\Gamma[\Upsilon \to e^+e^-]$ to obtain the ratio
\begin{equation}
R_0=\frac{\Gamma[\Upsilon \to \gamma\ A^0]}{\Gamma[\Upsilon \to e^+e^-]}=
\frac{m_{\Upsilon}^2\ X_d^2}
{8\pi\alpha\ v^2}\biggl(1-\frac{m_{A^0}^2}{m_{\Upsilon}^2}\biggr)
\label{eq:wtree}
\end{equation}
Strictly speaking, the above formula is only valid
when the photon energy is greater than the $\Upsilon$ binding energy. 
Actually, bound-state effects have a quite different behaviour
for a scalar or a pseudoscalar Higgs, reducing the former
substantially (even yielding a cancellation at a given
boson mass because
of the destructive interference of two amplitudes \cite{Faldt:1987zu}), 
but increasing $R_0$ by $\sim 20\%$ in the latter case
\cite{Polchinski:1984ag,Pantaleone:1984ug,Bernreuther:1985ja}. In addition,  
QCD corrections (at order $\alpha_s$) 
may reduce the ratio $R_0$ by a similar amount 
\cite{Vysotsky:1980cz,Nason:1986tr}. Thus 
the tree-level expression (\ref{eq:wtree}) can be used
for a pseudoscalar Higgs to an acceptable degree of accuracy.

On the other hand, the decay width of a CP-odd Higgs boson into
a tauonic pair is given by
\begin{equation}
\Gamma[A^0 \rightarrow \tau^+\tau^-] \simeq 
\frac{m_{\tau}^2X_d^2}{8\pi v^2}\ m_{A^0}\ (1-4x_{\tau})^{1/2}
\end{equation}
with $x_{\tau}=m_{\tau}^2/m_{A^0}^2$. Next, considering
the cascade decay $\Upsilon \to \gamma\ A^0(\to \tau^+\tau^-)$ 
shown in Fig.1.a), and
combining Eqs.(5-6) we can write  
\begin{equation}
{\cal R}_{\tau/\ell} = R_0
\times \frac{\Gamma[A^0 \rightarrow \tau^+\tau^-]}
{\Gamma_{A^0}}
\label{eq:ratiow}
\end{equation}
where $\Gamma_{A^0}$ denotes the full width of the pseudocalar 
Higgs boson.
Below open bottom production and above the $\tau^+\tau^-$
threshold, the tauonic channel should dominate the $A^0$ decay
even for moderate $X_d$, i.e. we can safely set
${\cal B}[A^0 \to \tau^+\tau^-] \approx 1$.
In section 4, $A^0-\eta_b$ mixing effects should modify
this assumption by changing both the total
and partial tauonic decay widths of the $A^0$ physical state.

\section{Resonant decay}
In Ref.~\cite{Sanchis-Lozano:2003ha}, use was made of time-ordered
perturbation theory to incorporate the effect of intermediate
$b\bar{b}$ bound states in the process of Eq.(2)
schematically shown in Fig.1.b). 
We found that the main contribution
should come from a $\eta_b$ state subsequent to a
dipole magnetic (M1) transition of the
$\Upsilon$ resonance, in agreement with \cite{Bernreuther:1985ja}.  
Thus, the total decay width
can be factorized in the narrow width approximation 
as \cite{Sanchis-Lozano:2003ha}
\footnote{A factorization
alternative to Eq.~(8) was also used in
\cite{Sanchis-Lozano:2003ha} 
based on the separation between
long- and short-distance physics following the main lines of
Non-Relativistic QCD \cite{Brambilla:2004wf} - albeit replacing a
gluon by a photon in the usual Fock decomposition of hadronic
bound states. Despite a (crude) estimate made in
\cite{Sanchis-Lozano:2003ha}, we find difficult to assess
the relevance of such a $|\eta_b^*+\gamma \rangle$ Fock component
and therefore we choose hereafter to base our estimates on 
factorization (\ref{eq:factor1}).}
\begin{equation}
\Gamma[\Upsilon\to\gamma \,\ell^+\ell^-]\ =\
\Gamma^{\,M1}_{\,\Upsilon\to\gamma \eta_b}\
\times\ \frac{\Gamma[\eta_b\to\ell^+\ell^-]}{\Gamma_{\eta_b}}
\label{eq:factor1}
\end{equation}
where $\Gamma[\eta_b\to\ell^+\ell^-]$ and
$\Gamma_{\eta_b}$ denote the leptonic width and the total width
of the $\eta_b$ resonance, respectively;
$\Gamma^{\,M1}_{\,\Upsilon\to\gamma \eta_b}$ represents
the M1 transition width. Next
dividing both sides of Eq.(\ref{eq:factor1}) by
the $\Upsilon$ total width $\Gamma_{\Upsilon}$, we get
the cascade decay formula
\[
{\cal B}[{\,\Upsilon\to\gamma \,\ell^+\ell^-}]=
{\cal B}[{\,\Upsilon\to\gamma \eta_b}] \times
{\cal B}[\eta_b \to \ell^+ \ell^-]
\]
The BF for a M1 transition between
$\Upsilon(nS)$ and $\eta_b(n')$ states ($n' \leq n$) is
given by \cite{Godfrey:2001eb}
\begin{equation}
{\cal B}[\Upsilon(nS) \to \gamma \eta_b(n'S)]
=\frac{16\alpha}{3}\biggl(\frac{Q_b}{2m_b}\biggr)^2\ 
\frac{I^2 \cdot k^3}{\Gamma_{\Upsilon}}
\end{equation}
where $k$ is the
photon energy (approximately equal to the mass difference
$m_{\Upsilon(nS)}-m_{\eta_b(n'S)}$);
$I$ represents the initial and
final wave functions overlap, $\langle f|j_0(kr/2)|i\rangle$, 
where $j_0$ is a spherical Bessel function.  
$I$ is numerically close to unity
for favored transitions ($n=n'$) but much smaller for
hindered ($n \neq n'$) transitions. 
As stressed in \cite{Godfrey:2001eb}, however, 
the considerably larger photon energy $k$ 
in the latter case could compensate 
this drawback, leading to competitive transition probabilities.
Therefore $\Upsilon(2S)$ and $\Upsilon(3S)$ hindered transitions
into $\eta_b(2S)$ and
$\eta_b(1S)$ states have to be taken into account as
potential contributions to the process of Eq.(2).

The $A^0$-mediated decay width of the $\eta_b$ into $\tau^+\tau^-$ 
can be related to the electromagnetic decay
of the Upsilon resonance as \cite{Sanchis-Lozano:2003ha}
\[
\frac{\Gamma[\eta_b \to \tau^+\tau^-]}{\Gamma[\Upsilon \to e^+e^-]}
\simeq \frac{3m_b^4m_{\tau}^2(1-4x_{\tau})^{1/2}X_d^4}{32\pi^2 Q_b^2 \alpha^2
\Delta M^2 v^4}
\]
where now $x_{\tau}=m_{\tau}^2/m_{\eta_b}^2$, 
$\Delta M=|m_{A^0}-m_{\eta_b}|$ and
the $A^0$ width was neglected. Note again that
the NP contribution to the $\Upsilon$ 
leptonic decay would be significant
only for the tauonic mode. Finally one gets for the ratio (\ref{eq:R})
\begin{equation}
{\cal R}_{\tau/\ell} \simeq \biggl[\frac{m_b^2m_{\tau}^2
(1-4x_{\tau})^{1/2}}
{8\pi^2{\alpha}v^4}\biggr]
\times \frac{I^2\ k^3X_d^4}{\Gamma_{\eta_b} {\Delta}M^2}
\label{eq:ratioeta}
\end{equation}

In order to crudely estimate $\Gamma_{\eta_b}$, 
we can employ the asymptotic pQCD expression \cite{oliver}
\[ 
\frac{\Gamma_{\eta_b}}{\Gamma_{\eta_c}} \simeq \frac{m_b}{m_c}
\times \biggl[\frac{\alpha_s(\mu_b)}{\alpha_s(\mu_c)}\biggr]^5 \]
Setting $m_b=4.7$ GeV, $m_c=1.5$ GeV, 
$\Gamma_{\eta_c}=25.5 \pm 3.4$ MeV \cite{Yao:2006px}
and varying the scales $\mu_c$ ($\mu_b$) 
between $m_c$ and $2m_c$ ($m_b$ and $2m_b$), one gets 
$\Gamma_{\eta_b} \simeq \Gamma[\eta_b \to gg] \sim 5-10$ MeV
although with a large uncertainty. Let us remark that a 
large $\eta_b$ hadronic decay width
(i.e. $\Gamma[\eta_b \to gg] > 10$ MeV) would damp
resonant effects in the process (2), 
as can be seen from Eq.(\ref{eq:factor1}).

For large enough $X_d$, the tauonic
decay mode of the $\eta_b$ becomes comparable 
to the hadronic mode provided by the 2 gluon channel
(even becoming dominant for $X_d \gtrsim 25$), i.e.
$\Gamma[\eta_b\to\tau^+\tau^-] \gtrsim \Gamma[{\eta_b} \to gg] \ $; thus
${\cal B}[\eta_b \to \tau^+\tau^-] \simeq 50-100 \%$ and 
\[ {\cal R}_{\tau/\ell} \simeq \frac{{\cal B}[\Upsilon\to\gamma \eta_b]}
{{\cal B}_{\ell\ell}}\ \simeq\ 1-10\ \% \]
using different values of ${\cal B}[\Upsilon\to\gamma \eta_b]$
summarized in Ref.~\cite{Godfrey:2001eb}. A similar result will
be obtained in the next section employing the mixing formalism.

\section{Mixing} So far we have neglected the widths of both
$\eta_b$ and $A^0$ in the theoretical 
calculation of the ratio (\ref{eq:R}).
However, for a light Higgs boson it may happen that
the $\eta_b$ and $A^0$ masses are close, e.g. within
several $\Gamma_{\eta_b}$, $\Gamma_{A^0}$. Then, 
a significant mixing between the
pseudoscalar resonance and a CP-odd Higgs boson should occur, 
modifying the naive estimates of (\ref{eq:ratiow}) and (\ref{eq:ratioeta}). 
Throughout we
will consider the simplified situation where the light
Higgs mixes appreciably with only one $\eta_b$ state,
notably the $\eta_b(1S)$ resonance.

The mixing between Higgs and resonances is
described by the introduction of off-diagonal
elements denoted by $\delta m^2$ in the mass matrix
\cite{Franzini:1985nt,Drees:1989du}
\[
{\cal M}_0^2=  
\left(
     \begin{array}{cc}
      m_{A_0^0}^2-im_{A_0^0}\Gamma_{A_0^0} & \delta m^2\\
      \delta m^2 & m_{\eta_{b0}}^2-im_{\eta_{b0}}\Gamma_{\eta_{b0}}      
     \end{array}
\right)
\] 
where the subindex $\lq$0' indicates unmixed states.
The off-diagonal element $\delta m^2$ 
can be computed (see Fig.1.c)
within the framework of a nonrelativistic quark potential model.
For the pseudoscalar case under study, one can write 
\begin{equation}
\delta m^2\ =\  
\biggl(\frac{3m_{\eta_{b}}^3}{4 \pi v^2}\biggr)^{1/2} |R_{\eta_b}(0)|
\times X_d\ 
\end{equation}
Notice that $\delta m^2$ is proportional to $X_d$; 
substituting numerical values (for the radial wave function at the origin
we use the potential model estimate 
$|R_{\eta_b}(0)|^2=$ 6.5 GeV$^3$ from \cite{Eichten:1995ch}) 
one finds:
$\delta m^2$(GeV$^2$) $\approx\ 0.146\ \times\ X_d$.

\begin{figure}
\begin{center}
\includegraphics[width=16pc]{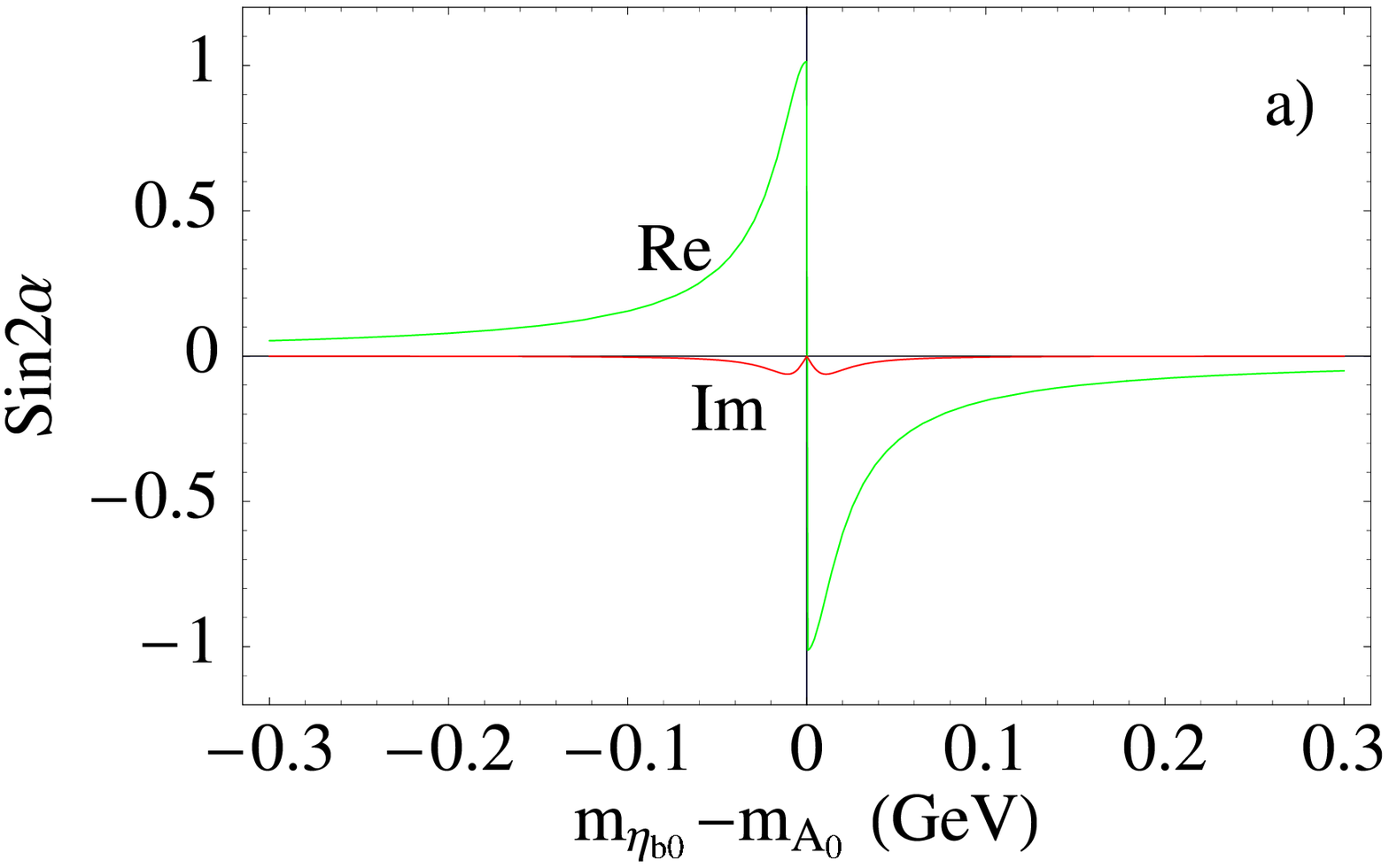}
\includegraphics[width=16pc]{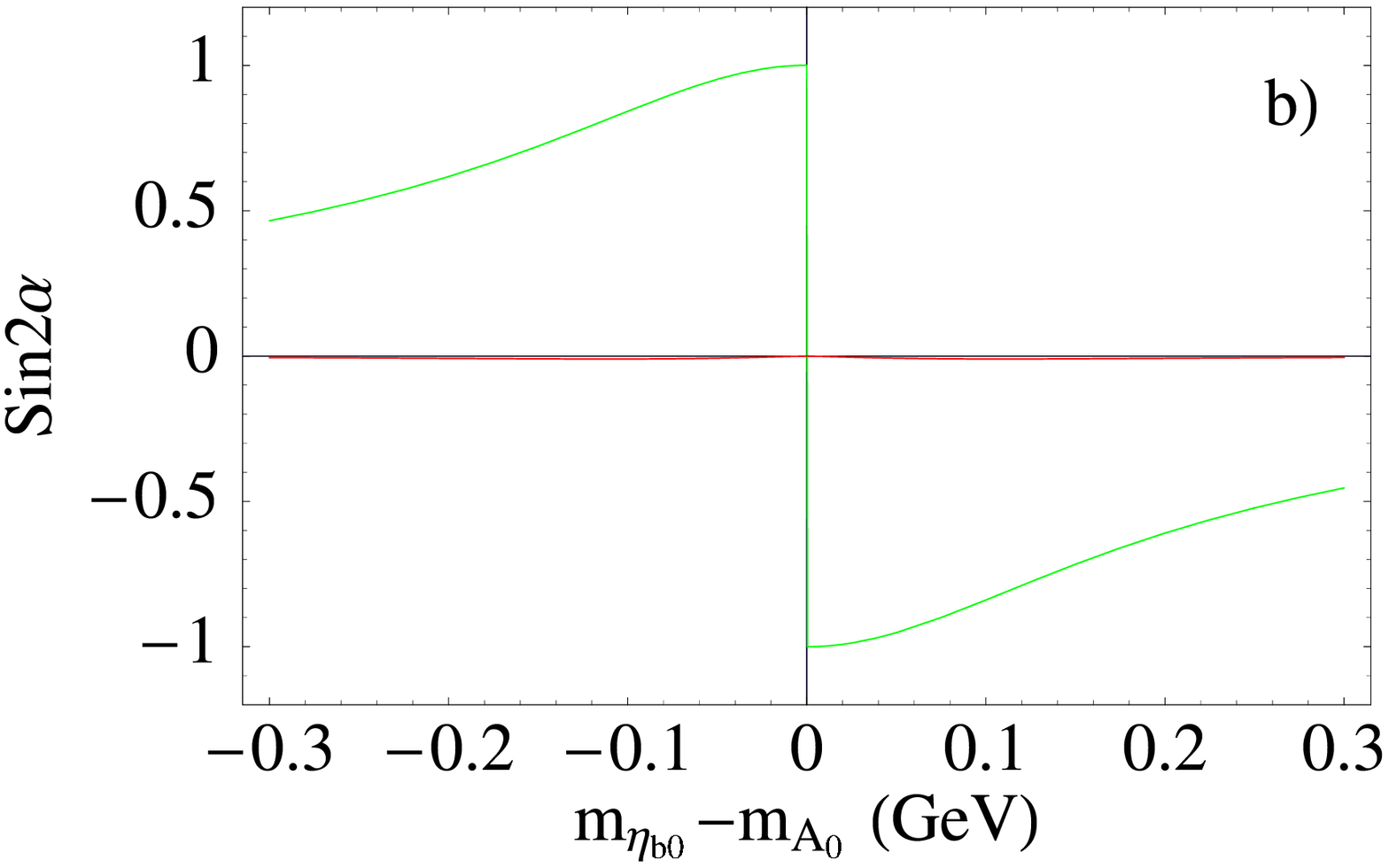}
\includegraphics[width=16pc]{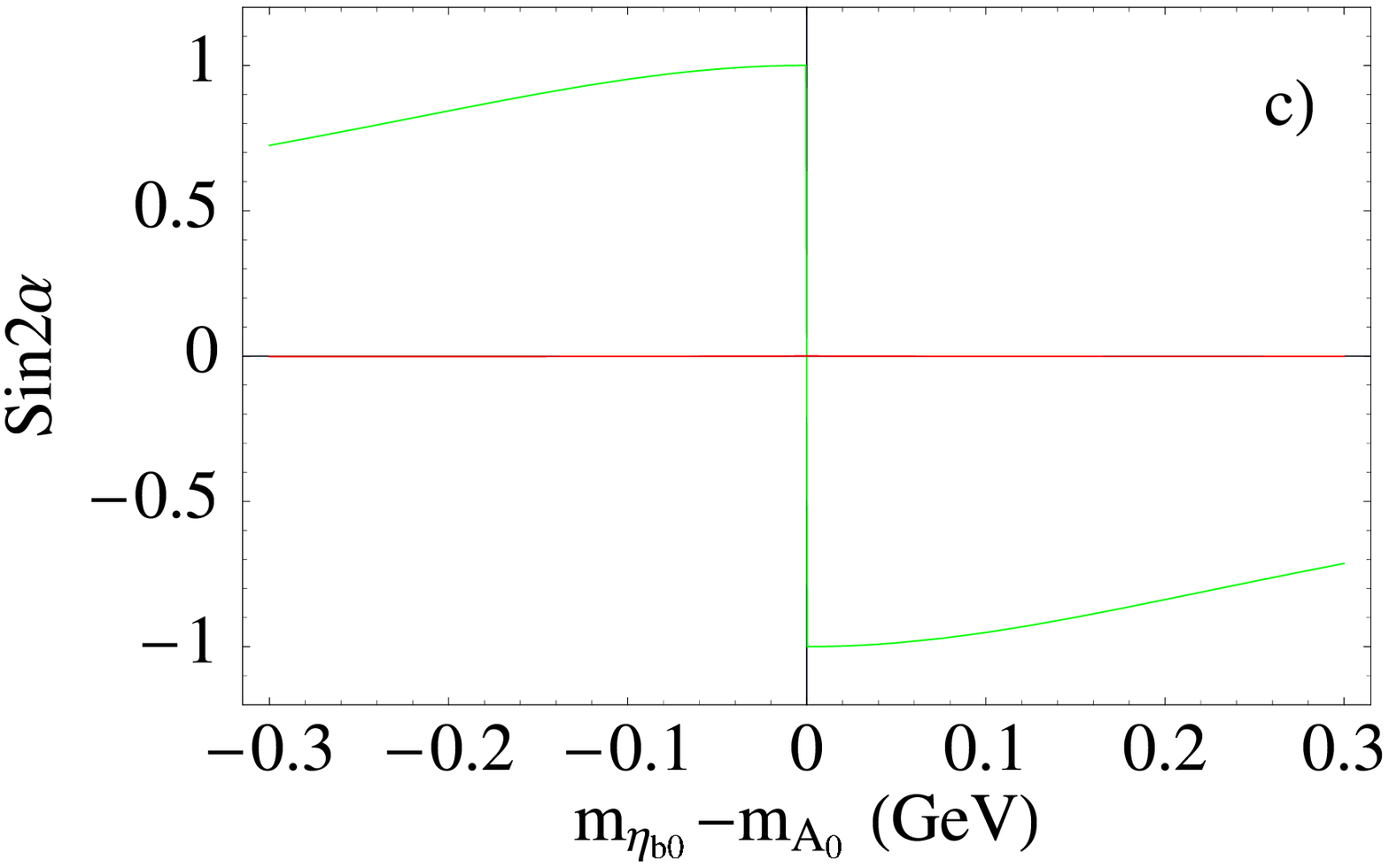}
\includegraphics[width=16pc]{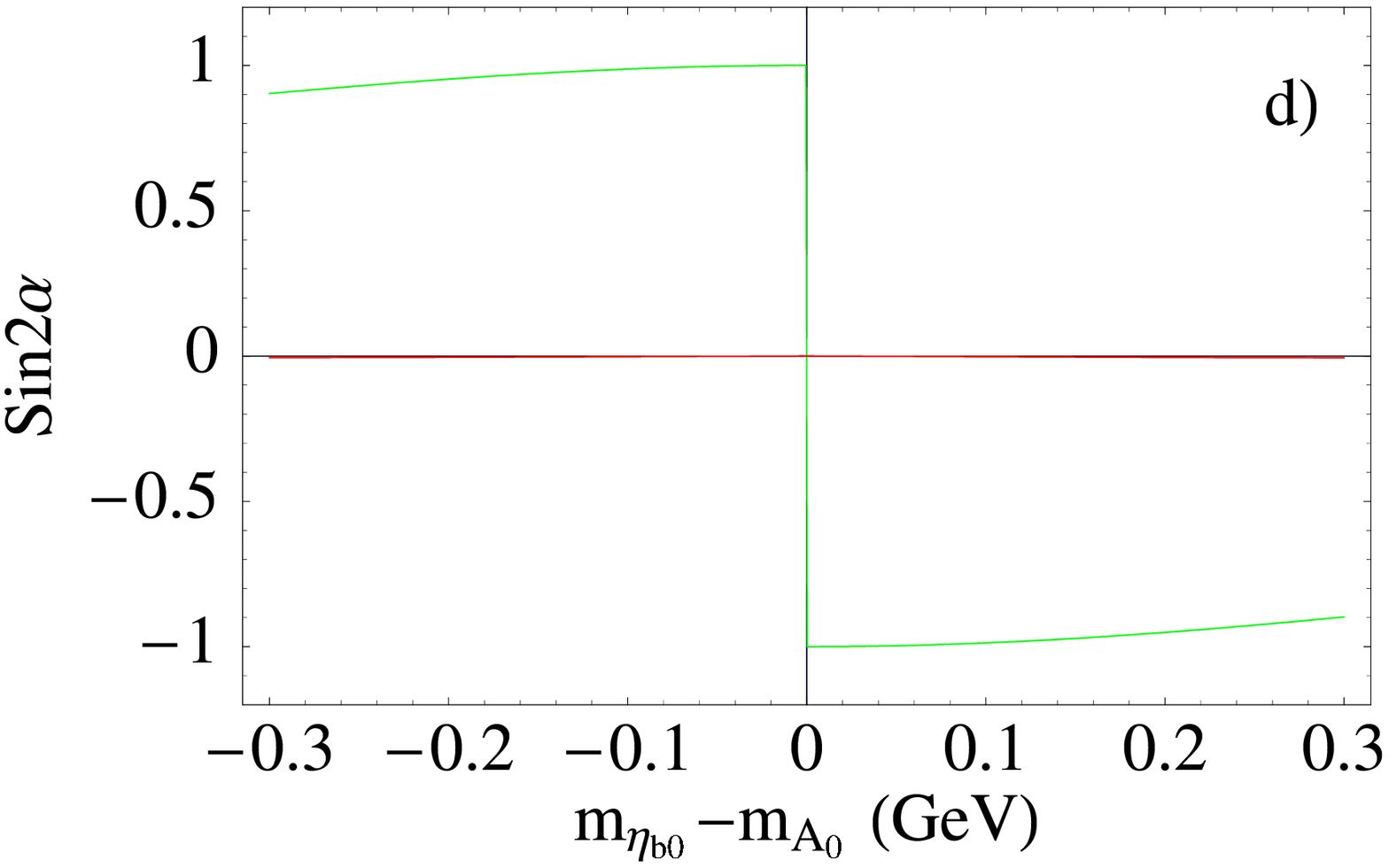}
\end{center}
\caption{$\sin{2\alpha}$ versus the mass difference of the unmixed states
setting $m_{\eta_{b0}}=9.3$ GeV and $\Gamma_{\eta_{b0}}=5$ MeV, 
for: $a)$ $X_d=1$ 
(Fig.2 of Ref.~\cite{Drees:1989du} is recovered); $b),c),d)$
$X_d=10,20,40$, respectively. Solid (green) line:
real part, dotted (red) line: imaginary part.
The change of sign of the real part is a consequence of the definition of the
mixing angle between the $A^0$ and $A_0^0$ states. The imaginary part
is negligible in our numerical computations and the normalization
$N$ of mixed states is near unity.}
\end{figure}

Now defining $\Delta^2= [D^2+(\delta m^2)^2]^{1/2}$, 
where $D=(m_{A_0^0}^2-m_{\eta_{b0}}^2-im_{A_0^0}\Gamma_{A_0^0}+
im_{\eta_{b0}}\Gamma_{\eta_{b0}})/2$, the mixing angle $\alpha$ 
between the physical Higgs $A^0$ and the unmixed $A_0^0$ 
can be defined through
\begin{equation}
\sin{}2\alpha\ =\ \delta m^2/\Delta^2
\end{equation}
giving rise to the
physical eigenstates:
\begin{eqnarray}
A^0 &=& \frac{1}{N}\{\cos{\alpha}\ A_0^0\ +\ \sin{\alpha}\ \eta_{b0}\} 
\nonumber \\
\eta_b &=& \frac{1}{N}\{\cos{\alpha}\ \eta_{b0}\ -\ \sin{\alpha}\ A_0^0\} 
\nonumber
\end{eqnarray}
where $N=\{|\cos{\alpha}|^2+|\sin{\alpha}|^2\}^{1/2}$, 
which can be different from unity since the mixing
angle is in general complex, although
we can set $N \simeq 1$ in our calculations.
 
Fig.3 shows $\sin{2\alpha}$
as a function of the $\eta_{b0}$ and $A_0^0$
mass difference for $X_d=1,10,20,40$ and $\Gamma_{\eta_{b0}}=5$ MeV.
Note that mixing effects can be important over a large mass
region (i.e. several hundreds of MeV), especially for large $X_d$ values. 

On the other hand, the couplings 
of the physical $A^0$ and $\eta_b$ states to a $\tau^+\tau^-$ pair 
are given by
\begin{eqnarray}
g_{A^0\tau\tau} &=& 
\frac{1}{N}\{\cos{\alpha}\ g_{A_0^0\tau\tau}\ +\ 
\sin{\alpha}\ g_{\eta_{b0}\tau\tau}\} 
\nonumber \\
g_{\eta_b\tau\tau} &=& \frac{1}{N}\{\cos{\alpha}\ g_{\eta_{b0}\tau\tau}\ -\ 
\sin{\alpha}\ g_{A_0^0\tau\tau}\} 
\nonumber
\end{eqnarray}

In the SM we can very approximately 
set $g_{\eta_{b0}\tau\tau} \simeq 0$ and 
thus write
\begin{equation}
g_{A^0\tau\tau} \simeq g_{A_0^0\tau\tau}\cos{\alpha},\ g_{\eta_b\tau\tau}
\simeq -g_{A_0^0\tau\tau}\sin{\alpha}
\end{equation}
where $g_{A_0^0\tau\tau}$ can be obtained from the Yukawa coupling strength
provided by Eq.(\ref{eq:yukawa}).
We should emphasize that, in order to account for
the decay $\Upsilon \to \gamma\ \tau^+\tau^-$,  
both channels represented in Eqs.(1) and (2) 
have to be considered altogether, taking into account the mixing diagram
of Fig.1.c) and modifying the tauonic couplings accordingly.

\begin{figure}
\begin{center}
\includegraphics[width=17pc]{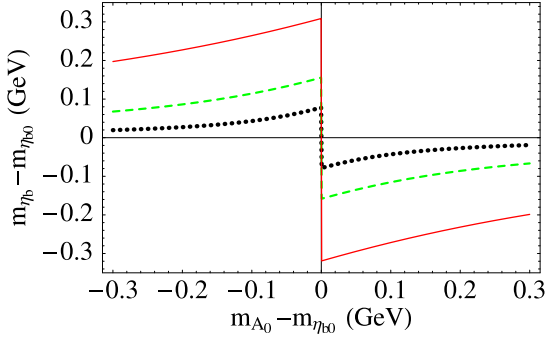}
\end{center}
\caption{Shift of the $\eta_b$ physical mass
induced by its mixing with a pseudoscalar Higgs boson, 
versus $m_{A_0^0}-m_{\eta_{b0}}$ for: 
dotted (black) line: $X_d=10$, dashed (green) line: $X_d=20$,
solid (red) line: $X_d=40$. The mass of the $\eta_b$ 
mixed state is decreased (increased) if $m_{A_0^0}>m_{\eta_{b0}}$
($m_{A_0^0}<m_{\eta_{b0}}$), ultimately implying a larger (smaller) 
$\Upsilon-\eta_b$ mass splitting than expected in the SM.} 
\end{figure}

Now, introducing the complex quantity
$E=m_{A_0^0}^2+m_{\eta_{b0}}^2-im_{\eta_{b0}}\Gamma_{\eta_{b0}}
-im_{A_0^0}\Gamma_{A_0^0}$, 
the masses and decay widths of the mixed 
(physical) states are given by
\begin{equation}
m_{1,2}^2-im_{1,2}\Gamma_{1,2}=
\frac{E}{2}\ {\pm}\ \Delta^2
\label{eq:mw}
\end{equation}
where subscripts $1,2$ refer to a Higgs-like state and a
resonance-like state respectively, if $m_{A_0^0}>m_{\eta_{b0}}$, 
and the converse if  $m_{A_0^0}<m_{\eta_{b0}}$. The real 
(imaginary) part of the r.h.s.
of Eq.(\ref{eq:mw})
provides the masses (widths) of the $A^0$ and $\eta_b$ 
physical states.

Moreover, the full widths $\Gamma_{A^0}$ and
$\Gamma_{\eta_b}$ of the physical states
can be expressed in terms of the widths of the unmixed states
according to the simple formulae:
\begin{eqnarray}
\Gamma_{A^0} &\simeq& |\cos{\alpha}|^2\ \Gamma_{A_0^0}\ +\  
|\sin{\alpha}|^2\ \Gamma_{\eta_{b0}}
\\
\Gamma_{\eta_b} &\simeq& |\cos{\alpha}|^2\ \Gamma_{\eta_{b0}}\ +\  
|\sin{\alpha}|^2\ \Gamma_{A_0^0}
\end{eqnarray}
Notice that $\Gamma_{\eta_{b0}}>>\Gamma_{A^0_0}$ was generally assumed
in Ref.~\cite{Drees:1989du}, while in this work
both $\eta_{b0}$ and $A_0^0$
widths can be comparable (the latter grows as $X_d^2$). In particular, 
for large $\delta m^2$, $\sin{2\alpha} \simeq 1$
at $m_{A_0^0}=m_{\eta_{b0}}$ and thus
$\Gamma_{\eta_{b}} \simeq \Gamma_{A^0} \simeq 
(\Gamma_{\eta_{b0}}+\Gamma_{A^0_0})/2$. 

In addition, spectroscopic effects can appear in 
$b\bar{b}(^1S_0)$ states if 
the $A^0_0-\eta_{b0}$ mixing sizeably shifts the
masses of the physical states. In Fig.~4 we plot
$m_{\eta_b(1S)}-m_{\eta_{b0}(1S)}$ versus 
$m_{A_0^0}-m_{\eta_{b0}}$. Such a shift has to be
added (with its sign) to the QCD expected 
$m_{\Upsilon(1S)}-m_{\eta_b(1S)}$ hyperfine splitting,
whose theoretical prediction is achieving
a remarkable precision \cite{Brambilla:2004wf}. 
As a consequence, if $m_{\eta_{b0}}<m_{A_0^0}$,  
the  hyperfine splitting could considerably increase 
with respect to the SM expectations even at moderate $X_d$ values. 
Let us remark that, in this case,   
the energy $k$ of the radiated photon could reach values well above 100 MeV
in the $\Upsilon(1S) \to \eta_b(1S)$ M1 transition, substantially
raising the rate due to the $k^3$ dependence.

\begin{figure}
\begin{center}
\includegraphics[width=19pc]{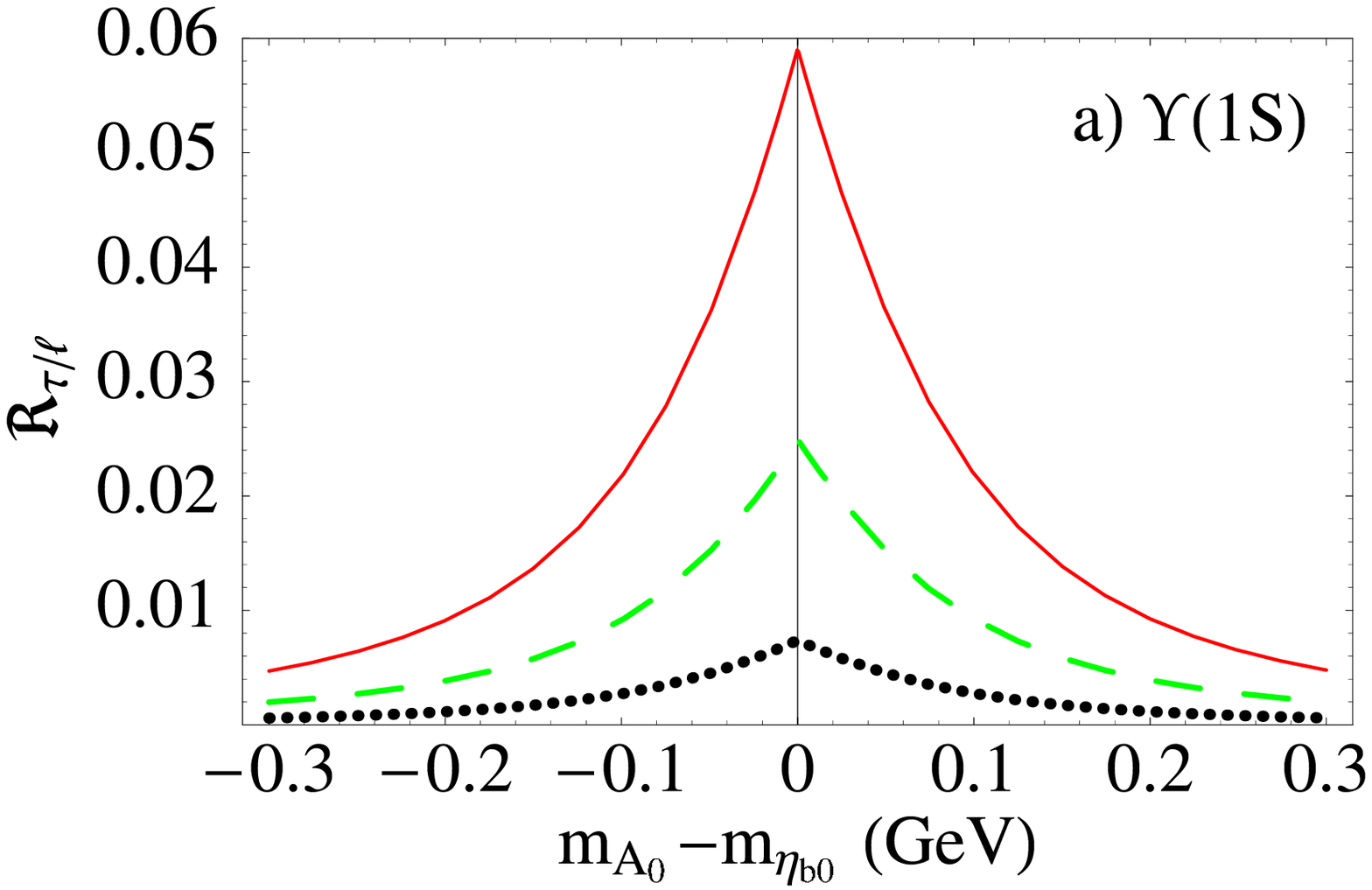}
\includegraphics[width=19pc]{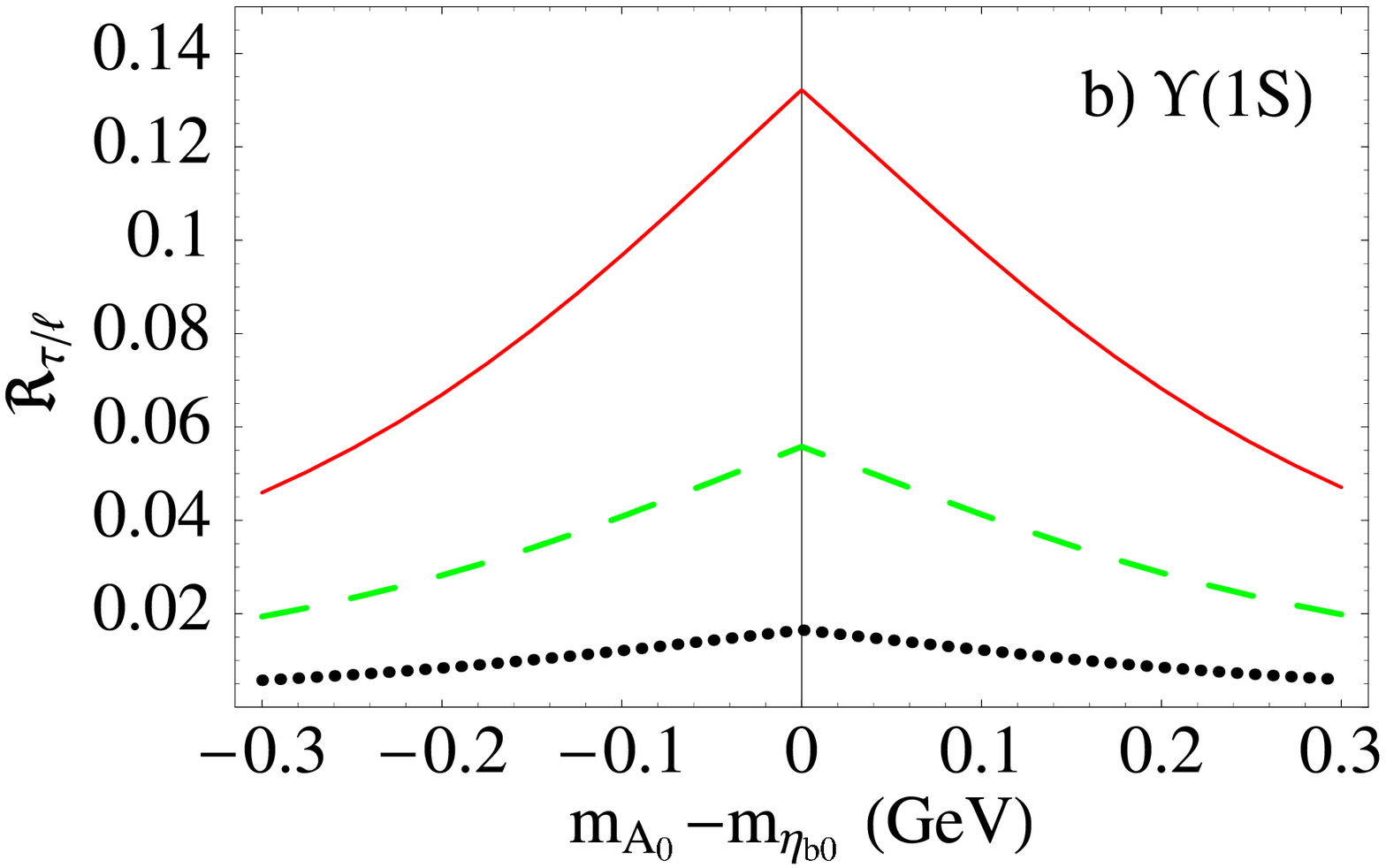}
\includegraphics[width=19pc]{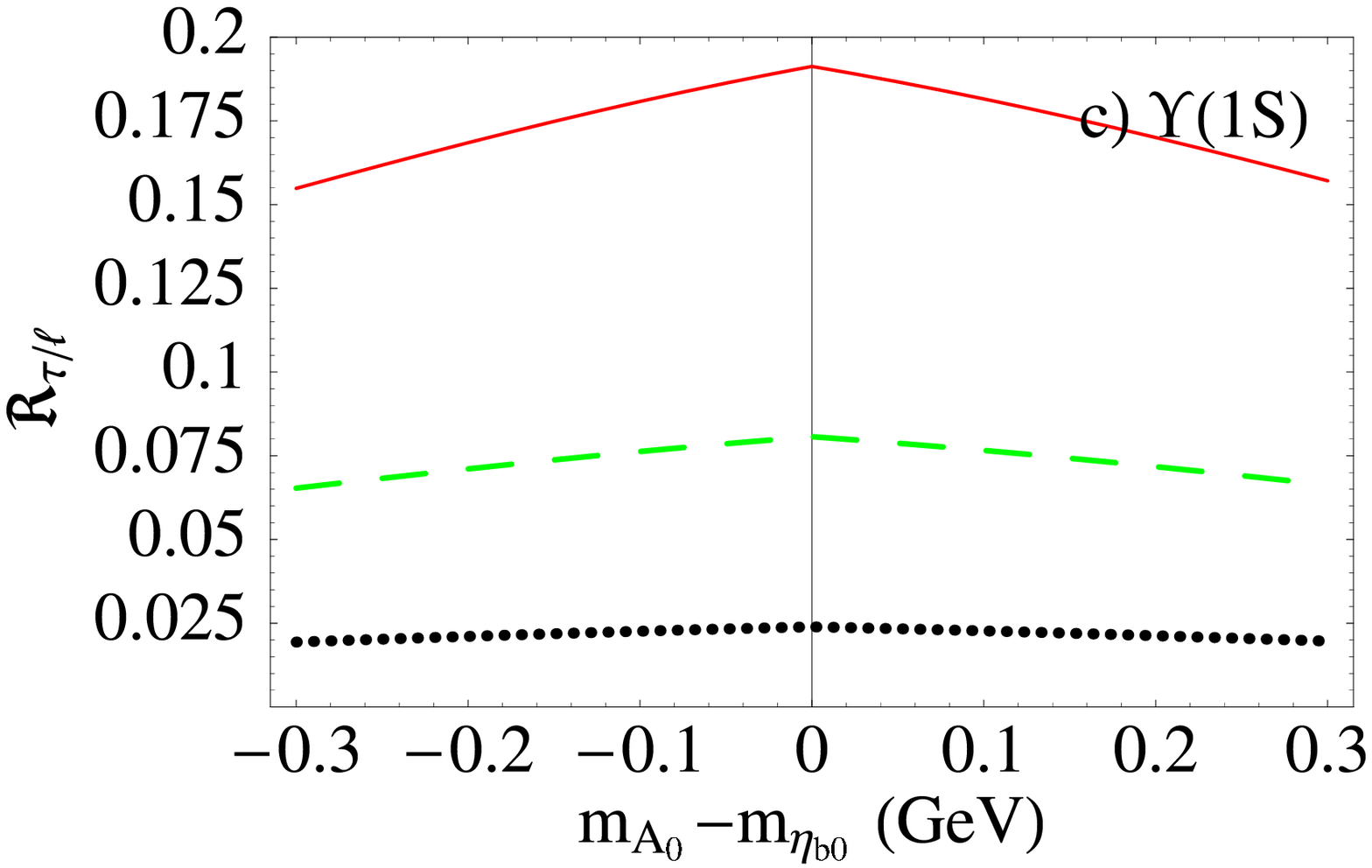}
\caption{$R_{\tau/\ell}$ from Eq.(18) (resonant production)
versus the mass difference
of the unmixed states for: $\Upsilon(1)$ with $X_d=$
$a)$ 10, $b)$ 20, $c)$ 40, respectively. In each plot,
$k=100$ MeV (dotted black line), $k=150$ MeV (dashed green line) and
$k=200$ MeV (solid red line).}
\end{center}
\end{figure}

\begin{figure}
\begin{center}
\includegraphics[width=19pc]{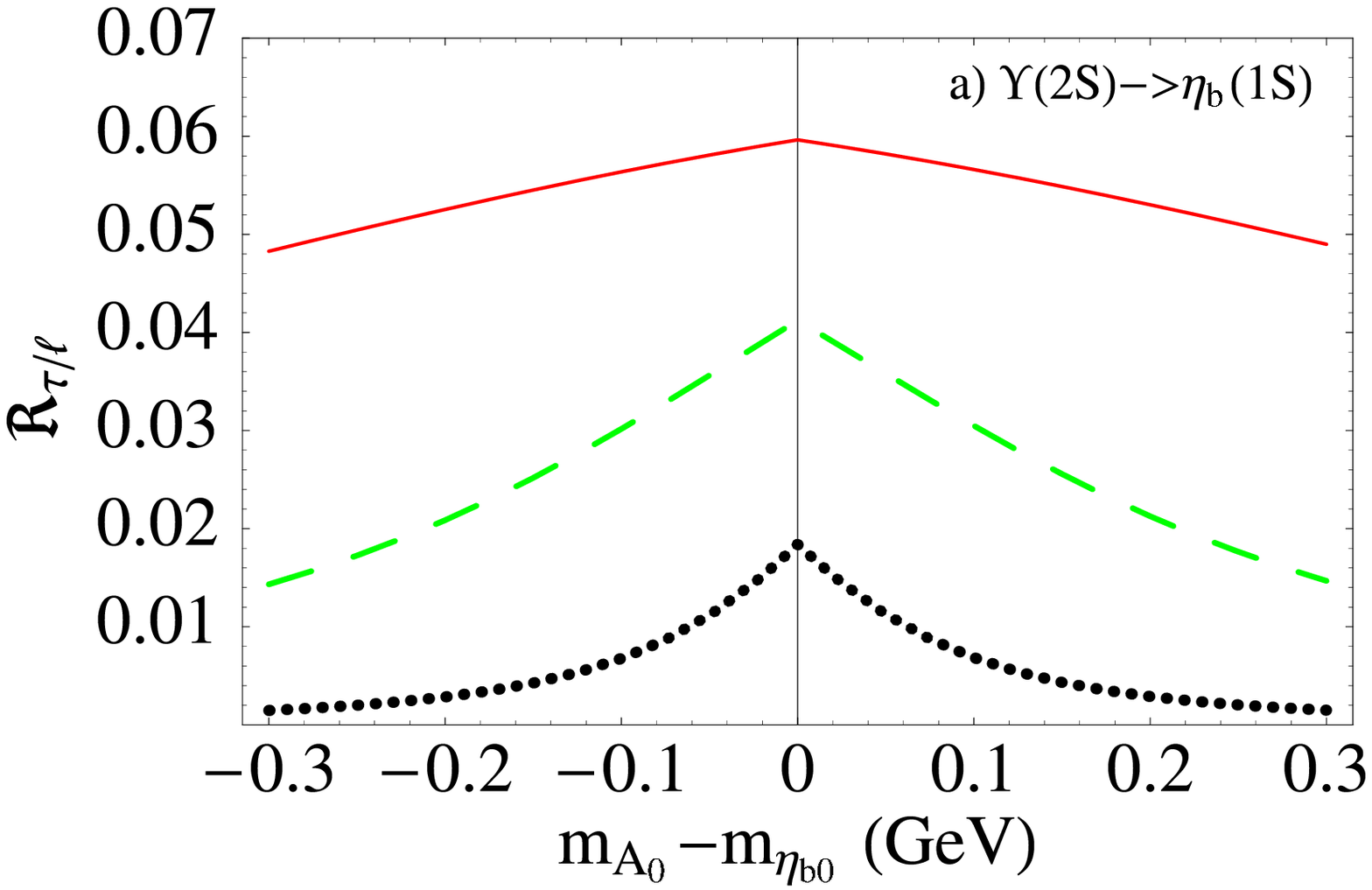}
\includegraphics[width=19pc]{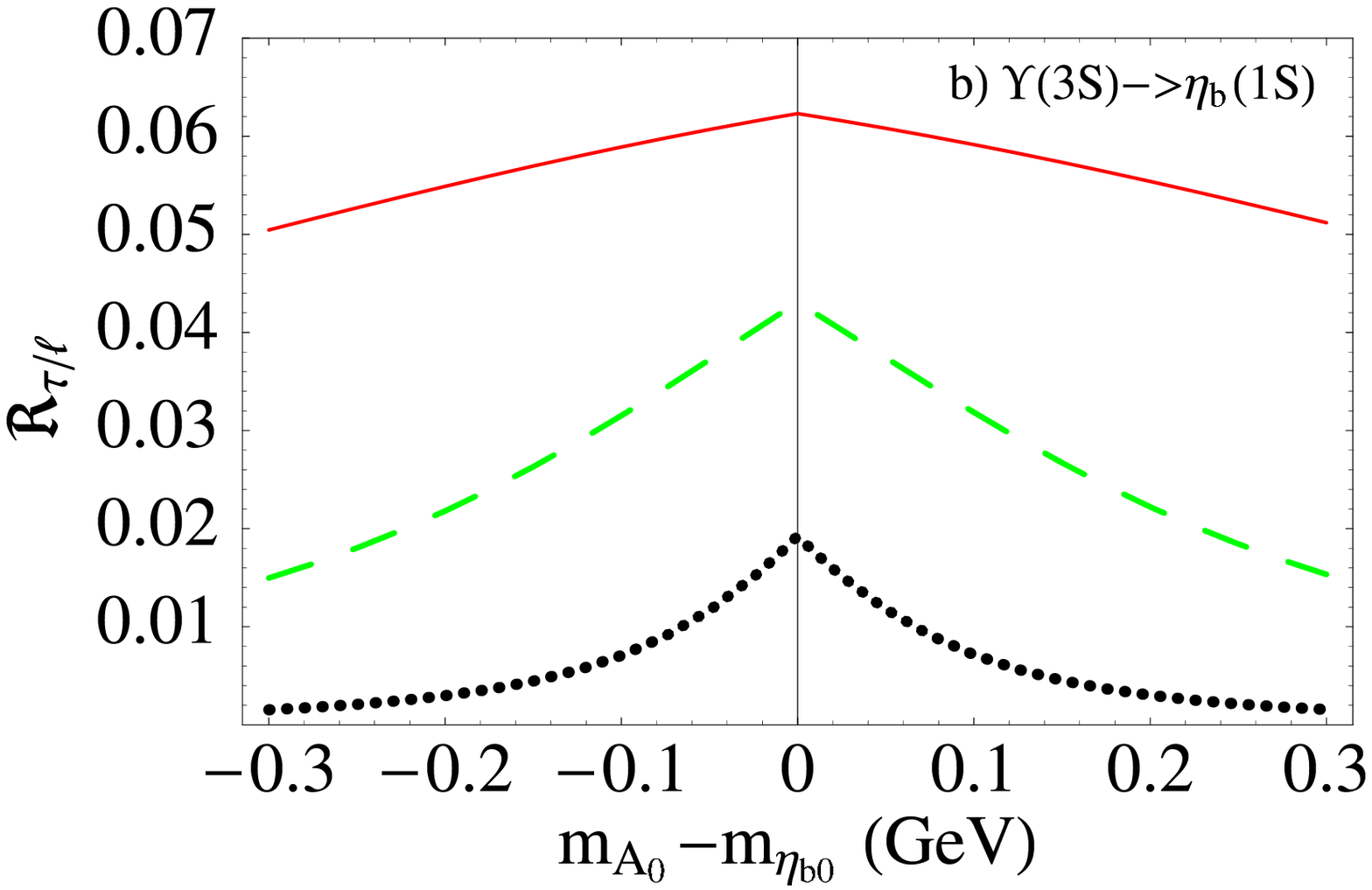}
\includegraphics[width=19pc]{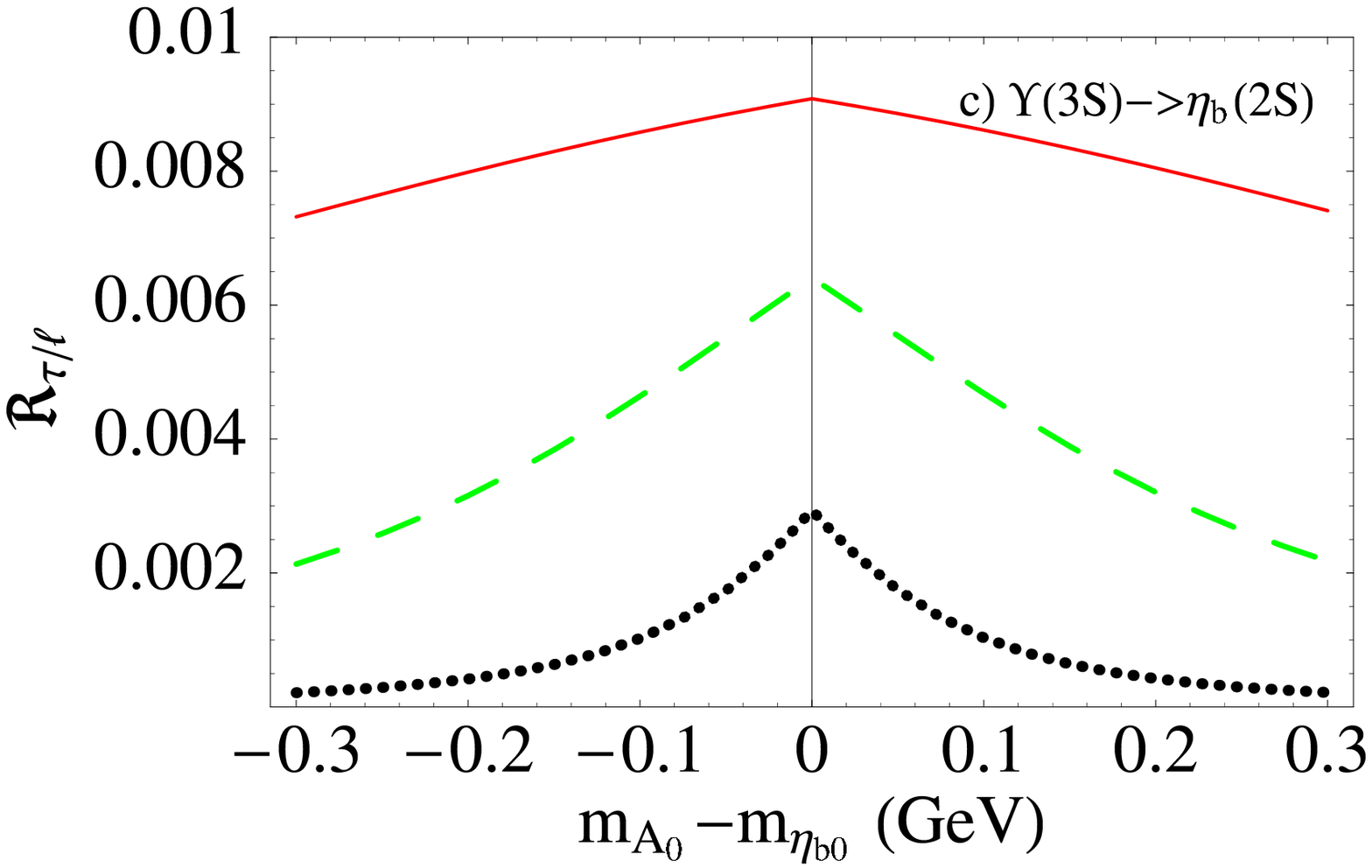}
\caption{$R_{\tau/\ell}$ from Eq.(18) (resonant production) 
versus the mass difference 
of the unmixed states for: 
$a)$  $\Upsilon(2S) \to \gamma \eta_b(1S)$, 
$b)$ $\Upsilon(3S) \to \gamma \eta_b(1S) $,   
and c) $\Upsilon(3S) \to \gamma \eta_b(2S)$, respectively.
We set $m_{\eta_b(1S,2S)}=9.3$, 9.96 GeV,   
$X_d=10$ (dotted black line), 20 (dashed green line) and 
40 (solid red line).}
\end{center}
\end{figure}

\section{Numerical Analysis}

The ultimate goal of this study is to evaluate 
the level of LU breaking in $\Upsilon$ decays
due to the NP contribution 
when mixing effects are taken into account.
To this end, we ought to modify the expressions (7) and (10)
of $R_{\tau/\ell}$
corresponding to non-resonant and resonant contributions
to the decay, by changing the widths in accordance with Eqs.(15-16).
Hence Eq.(7) becomes
\begin{equation}
{\cal R}_{\tau/\ell} = R_0
\times \frac{|\cos{\alpha}|^2\ \Gamma[A_0^0 \rightarrow \tau^+\tau^-]}
{|\cos{\alpha}|^2\Gamma_{A_0^0}+  
|\sin{\alpha}|^2\Gamma_{\eta_{b0}}}
\label{eq:ratiow2}
\end{equation}
where $\alpha$ depends on the masses of the
unmixed states $A_0^0$ and $\eta_{b0}$ following (12).
The corresponding formula for resonant production using the mixing
formalism can be written as
\begin{equation}
R_{\tau/\ell}=\frac{{\cal B}[\Upsilon \to \gamma \eta_b]}
{{\cal B}_{\ell\ell}} \cdot
\frac{|\sin{\alpha}|^2\Gamma[A_0^0 \to \tau^+\tau^-]}
{|\cos{\alpha}|^2\Gamma_{\eta_{b0}}+|\sin{\alpha}|^2\Gamma_{A_0^0}}
\end{equation}
The above expression results in Eq.(\ref{eq:ratioeta})
for small mixing angles, as can be verified 
using Eqs.(9-12) for
$|m_{\eta_b}-m_{A^0}| > \Gamma_{A^0}, \Gamma_{\eta_b}$,  
under the approximation $m_{\Upsilon} \simeq m_{\eta_b} \simeq 2m_b$.

From an experimental viewpoint, both
statistical and systematic  
uncertainties in the $R_{\tau/\ell}$
measurement
were estimated to be of order of 
the few percent level in a proposal  
for testing LU at a (Super) B factory 
\cite{Sanchis-Lozano:2006gx}. Due to the 
high-luminosity of the machine, the actual limitation for 
the measurement accuracy
should come from the systematic error,
expected to be $\lesssim 3.7\%$, representing a reference value for
the observability of NP effects in LU analyses.

In Fig.5 we plot $R_{\tau/\ell}$ 
versus $m_{A_0^0}-m_{\eta_{b0}}$ 
for the $\Upsilon(1S)$ resonant decay
via an intermediate $\eta_b(1S)$ state, using $X_d=10,20,40$
and $k=100,150,200$ MeV. Observable effects
only become likely for $X_d \gtrsim 10$ and
$k \gtrsim 150 $ MeV. This requirement implies
a hyperfine splitting $\Upsilon(1S)$-$\eta_b(1S)$
larger than expected in the SM but made possible by the 
mixing as shown in Fig.4.

Figs. 6.a) and b) show 
$R_{\tau/\ell}$ for $\Upsilon(2S,3S)$ resonant decays
via an intermediate $\eta_b(1S)$ state, 
versus $m_{A_0^0}-m_{\eta_{b0}}$ for $X_d=10,20,40$, setting
$I=0.057,0.029$ as overlap factors 
respectively \cite{Godfrey:2001eb}. 
In Fig.6.c) we plot $R_{\tau/\ell}$
for $\Upsilon(3S)$ resonant decays via a $\eta_b(2S)$
state for $k=400$ MeV and $I=0.054$ \cite{Godfrey:2001eb}.
The latter decay is significantly more suppressed than the 
former two and will be ignored. A caveat is in order. 
All these rates depend
strongly on $I$ and $\Gamma_{\eta_{b0}}$,
both quite uncertain.

\begin{figure}
\begin{center}
\includegraphics[width=19pc]{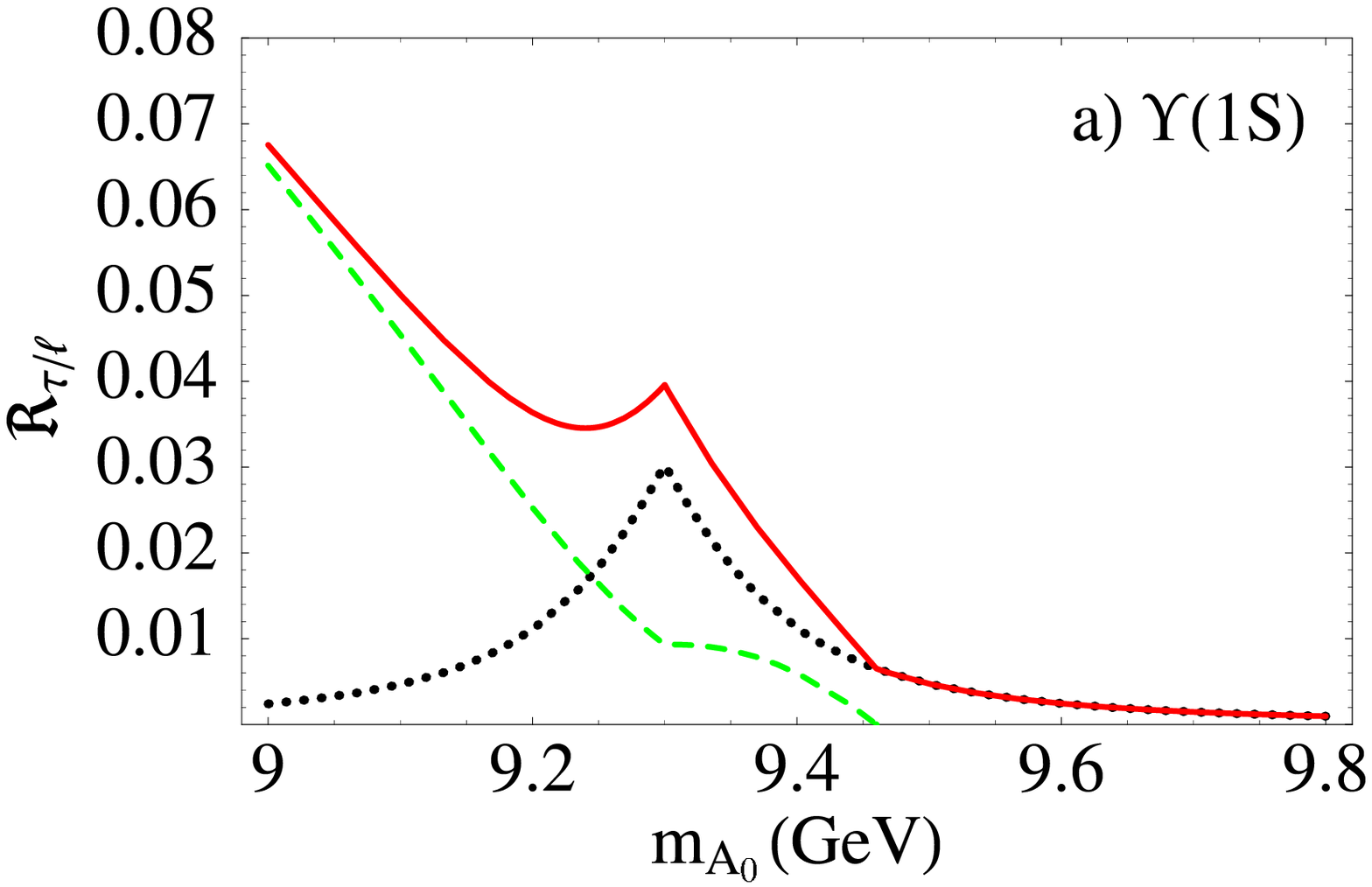}
\includegraphics[width=19pc]{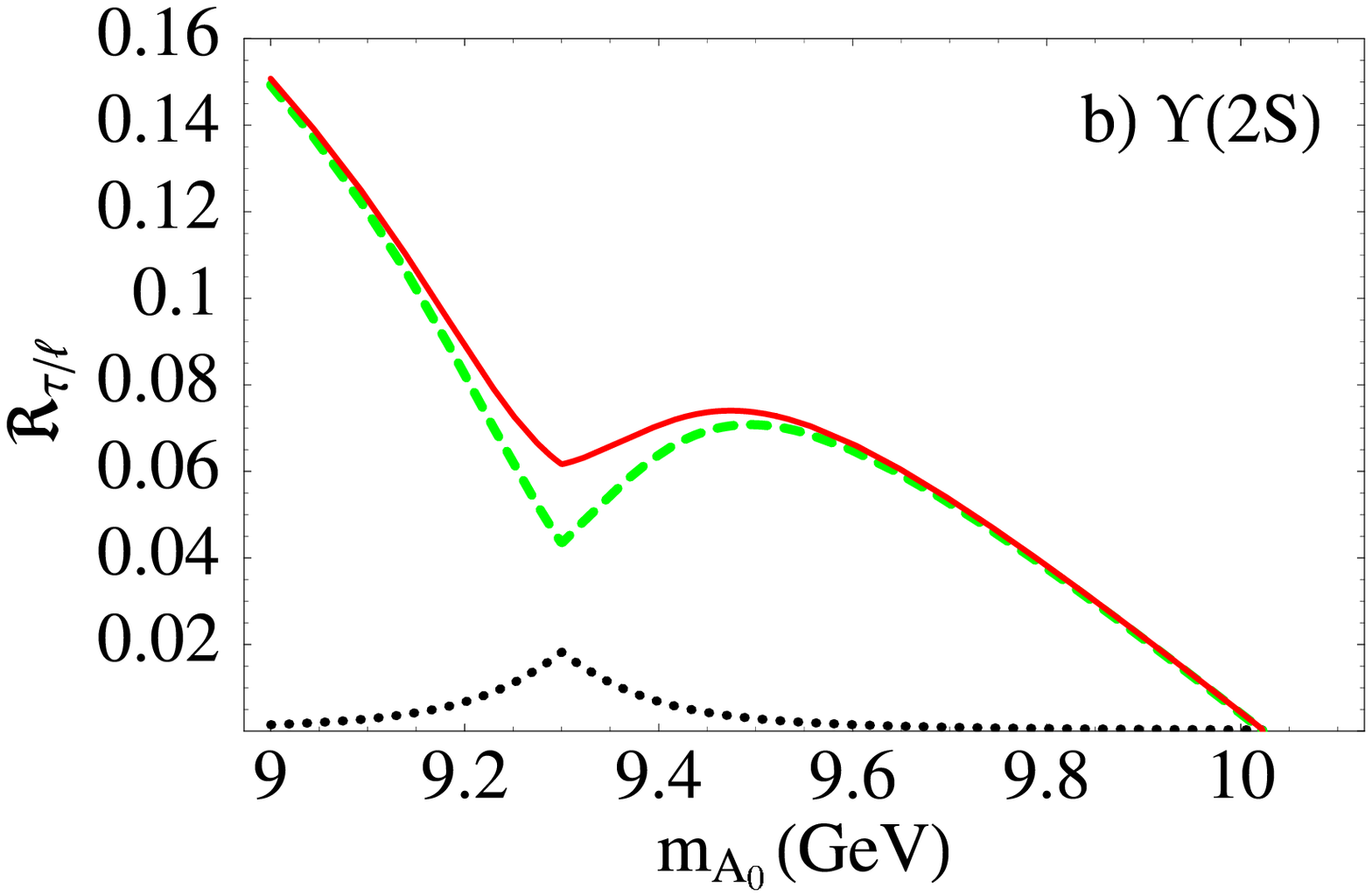}
\includegraphics[width=19pc]{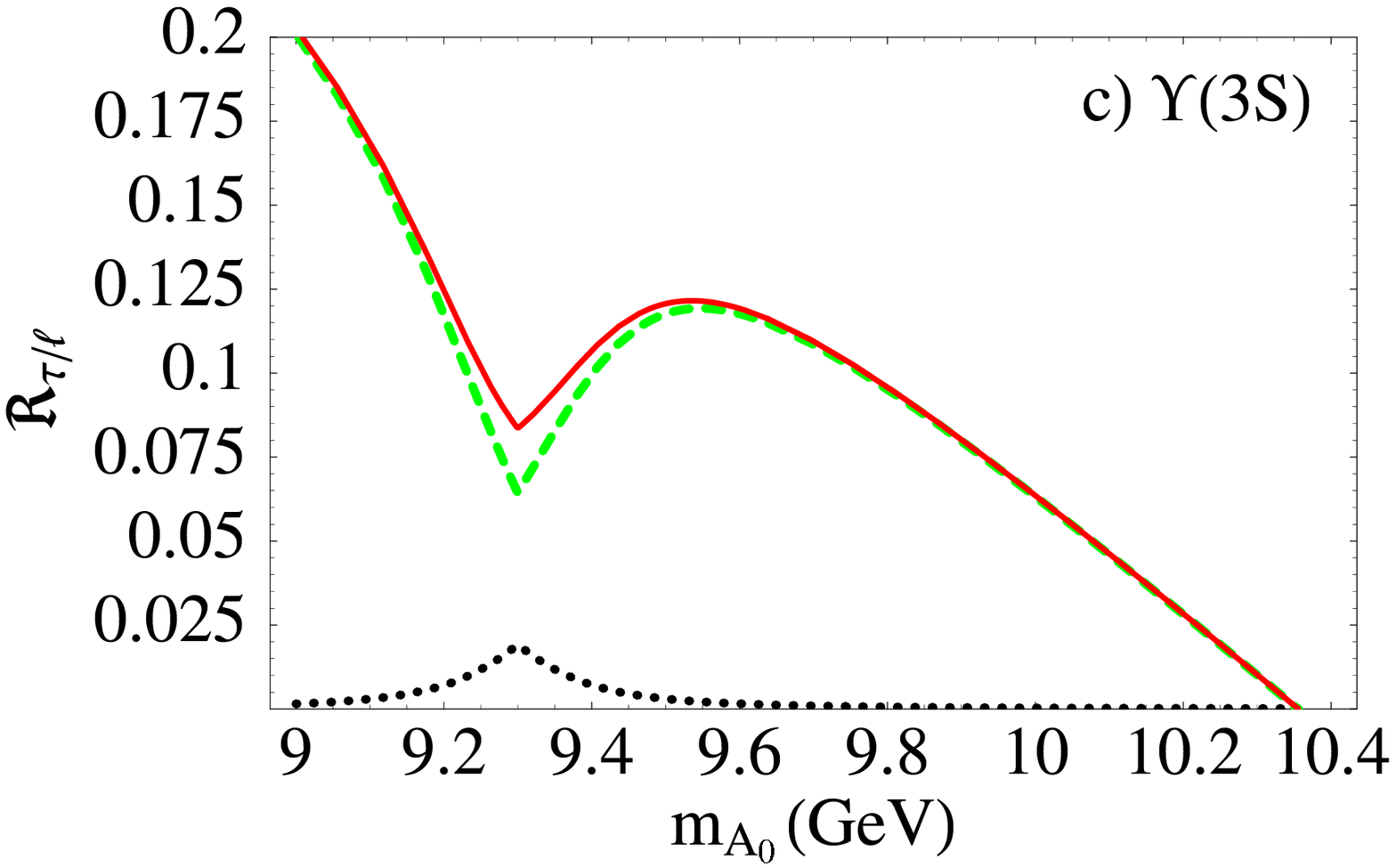}
\end{center}
\caption{$R_{\tau/\ell}$ versus the pseudoscalar Higgs mass
for $a)$ $\Upsilon(1S)$,  
$b)$ $\Upsilon(2S)$, and $c)$ $\Upsilon(3S)$ decays 
using $X_d=10$, $m_{\eta_{b0}}=9.3$ GeV, 
and $\Gamma_{\eta_{b0}}=5$ MeV, respectively. Resonant 
(dotted black line) and non-resonant (dashed green line)
decays are added in the solid red line. The non-resonant contribution
is shaped by the $|\cos{\alpha}|^2$ dependence
of the $A^0$ partial decay width in Eq.(17), due to the mixing with the 
pseudoscalar resonance, and the variation of the $A^0$
full width. Larger (smaller) values of $X_d$ obviously yield higher
(lower) expectations for $R_{\tau/\ell}$.}
\end{figure}

In Figures 7 we plot altogether resonant 
and non-resonant decays 
for the $\Upsilon(1S,2S,3S)$ resonances
with $\Gamma_{\eta_{b0}}$=5 MeV and $X_d$=10 as reference values. 
According to the NMSSM, 
$X_d=\cos{\theta_A} \tan{\beta}$, where $\cos{\theta_A}$
is expected typically small \cite{Dermisek:2006py}, i.e. 
$\cos{\theta_A} \sim few \times 10^{-1}$. Therefore
$X_d=10$ implies     
$\tan{\beta} \sim several \times 10$. Furthermore, note that in models 
other than the NMSSM, 
the singlet $A^0$ component might not need to be so 
large and, consequently,
$\cos{\theta_A}$ could be not so small, making 
larger $X_d$ plausible while keeping reasonable $\tan{\beta}$ values.

The set of plots of Fig.7 constitute our main result from which
we draw the final conclusions.
By inspection, a bump can be seen in Fig.7.a) due to the
resonant contribution, while a dip appears in Figs.7.b) and c) 
on account of the suppressed non-resonant channel, not
compensated by the resonant channel. In spite of that, the
higher values of $R_{\tau/\ell}$ obtained
in the two latter cases allow us to conclude that
$\Upsilon(2S,3S)$ radiative decays look more promising than
the $\Upsilon(1S)$ decays for the experimental observation 
of LU breaking at the few percent level.
This conclusion is important if
a specific test of LU were to be put forward by
experimental collaborations.

\section{Summary}

Lepton universality
breaking in $\Upsilon$ decays
could be a harbinger of new physics,
in particular of a light CP-odd 
non-standard Higgs boson.
In this paper, we have studied non-resonant and resonant 
$\Upsilon$ decays taking into account
$A^0-\eta_b$ mixing effects, 
concluding that, for this particular goal of
testing LU, 
$\Upsilon(2S,3S)$ radiative decays look more
promising than the $\Upsilon(1S)$ decay. Moreover,  
the discovery of $\eta_b$ states 
and a light CP-odd Higgs boson would become both experimentally
and theoretically entangled.
Indeed, the observation of a large tauonic BF of any $\eta_b$ state 
could be related to the existence of a light Higgs boson. In addition, 
unexpectedly large hyperfine splittings in the bottomonium family 
would also be another hint of new physics. 

High luminosity (Super) B factories should play 
a crucial role for testing LU in $\Upsilon$ decays
(perhaps below the few percent level if systematics are
well under control) in the quest for new physics at this 
energy scale. 

{\em Acknowledgements}
We gratefully acknowledge J.~Bernabeu, N.~Brambilla, F.J.~Botella,
J.~Papavassiliou and A.~Vairo for
useful discussions. M.A.S.L. thanks R.~Dermisek, 
G.~Faldt, J.~Gunion, B.~McElrath, P.~Osland and T.~T.~Wu for 
valuable exchange of emails.

\end{document}